\documentclass{pasj00}

\begin{document}
\SetRunningHead{Morita et al.}{T Tauri stars in L1014}
\Received{2006/07/18}
\Accepted{2006/08/04}

\title{Probable association of T Tauri stars with the L1014 dense core}

%
 \author{%
 Atsuko \textsc{Morita},\altaffilmark{1}
 Makoto \textsc{Watanabe},\altaffilmark{2}
 Koji \textsc{Sugitani},\altaffilmark{1}
 Yoichi \textsc{Itoh},\altaffilmark{3}
 Mariko \textsc{Uehara},\altaffilmark{4}\\
 Chie \textsc{Nagashima},\altaffilmark{4}
 Noboru \textsc{Ebizuka},\altaffilmark{5, 7}
Takashi \textsc{Hasegawa},\altaffilmark{6}
 Kenzo \textsc{Kinugasa},\altaffilmark{6}\\
   and
  Motohide \textsc{Tamura}\altaffilmark{7}}
 \altaffiltext{1}{Graduate School of Natural Sciences, Nagoya City University, Nagoya 467-8501, Japan}
 \altaffiltext{2}{Subaru Telescope, National Astronomical Observatory of Japan, Hilo, HI 96720, USA}
 \altaffiltext{3}{Graduate School of Science and Technology, Kobe University, Kobe 657-8501, Japan}
 \altaffiltext{4}{Department of Astrophysics, Nagoya University, Nagoya 464-8602, Japan}
 \altaffiltext{5}{RIKEN, Wako, Saitama 351-0198, Japan}
\altaffiltext{6}{Gunma Astronomical Observatory, Takayama-mura Agatsuma-gun Gunma 377-0702, Japan}
 \altaffiltext{7}{National Astronomical Observatory of Japan, Mitaka, Tokyo 181-8588, Japan}
 

\maketitle

\begin{abstract}
Using the Wide Field Grism Spectrograph 2 (WFGS2), we have carried out slit-less spectroscopy, 
$g'r'i'$ photometry, and slit spectroscopy on the L1014 dense core.  We detected three H$\alpha$ 
emission line stars.  We interpret one as weak-line T Tauri star (WTTS) and the others as 
classical T Tauri stars (CTTS).
Since their $g'-i'$ colors and/or classified spectral types are consistent with those of T Tauri stars 
and two of them show less extinction than the cloud, these three stars are likely to be T Tauri stars 
associated with L1014.
Adopting an age range for T Tauri stars, 1--10 Myr, the  color-magnitude diagram suggests 
a  distance of  $\sim$ 400--900 pc, rather than the previously assumed distance, 200 pc.
This could strongly affect on the mass estimate  of L1014-IRS, which is thought to be 
either a very young protostar or proto-brown dwarf.
\end{abstract}

\section{Introduction}

In the early evolution of substellar objects, the formation of brown dwarfs 
in molecular clouds is of particular interest; several formation mechanisms 
have been proposed (e.g., \cite{ML06, luhman06, whitworth06}). 
One possibility is the same mechanism as low-mass stars, 
but with much smaller cores (e.g., \cite{PN04}).
The L1014 dense core is a potential site for such a process.

L1014 was previously thought to be starless because it lacks
IRAS point sources  \citep{LM99}.   \citet{yng04} discovered 
a very faint infrared source (L1014-IRS) toward the center of L1014 
with the Spitzer Space Telescope.
Recent detections of a molecular outflow and a reflection nebula revealed 
that this central object is really embedded within the cloud core and is not 
a background source associated with the Perseus arm at 2.6 kpc 
\citep{bourke05, huard06}.
Assuming the distance of L1014 is $\sim$ 200 pc, as adopted in recent literature, 
this embedded source becomes an ideal candidate for a very low mass protostar, 
i.e., a proto-brown dwarf.

However, the mass estimate of a protostellar object depends on its adopted distance.
It is, in general,  not easy to determine the distance to  the surrounding dark cloud, 
particularly  that of a small, nearby dense core as  its small size implies 
a small number of background/foreground stars.
In fact, the assumption that L1014 is 200 pc away seems to originate from 
the distance estimate  of B362, which is located $\sim10\arcmin$  
north of L1014, but this estimate is based upon the similarity of $V_{LSR}$
 \citep{dieter1973} to that of nearby clouds with a kinematic distance 
 $\sim$ 100--200 pc \citep{dickman1976}.

The L1014 dense core is thought to be a site of very  recent  ($<$ 1 Myr) star formation.
We interpret the slightly extinct region extending southwest of L1014 in the DSS-II red 
image as thinner molecular gas extending from the dense core.  In fact, CO observations 
\citep{RP92, crapsi05} showed the existence of molecular material in this region.
We suspected that the thin molecular gas and the dense core are remnants 
of previous low-mass star formation over the past few Myr.
If so, pre-main-sequence stars, i.e., T Tauri stars,  could be detected toward L1014 
and the validity of the distance estimate of L1014 can be examined with the color-magnitude 
diagram by using theoretical isochrones of pre-main-sequence stars.

In order to substantiate the presence of previous star formation, we surveyed 
H$\alpha$ emission line stars by slit-less spectroscopy around this remarkable core.
Slit spectroscopy and $g'r'i'$ photometry were then performed on the selected stars. 
In this letter, we present our spectroscopic results and discuss whether 
the detected H$\alpha$ emission line stars are T Tauri stars associated with the L1014 core. 
We also discuss the distance to L1014. 

\section{Observations and Results}

Slit-less grism spectroscopy and $g'r'i'$ photometry of L1014 were made with 
the Wide Field Grism Spectrograph 2 (WFGS2; \cite{uehara04}) mounted on the
University of Hawaii (UH) 2.2-m telescope on 2004 November 13 UT.
The detector used was a Tektronix 2048$\times$2048 CCD.
The field of view is 11$\farcm5 \times 11\farcm$5 with a pixel scale of
0$\farcs$34 pixel$^{-1}$.
For the slit-less spectroscopy, we took three 300 s exposures dithered by
$\sim$ 20$\arcsec$ with a wide H$\alpha$ filter (FWHM = 50 nm) and
a 300 line mm$^{-1}$ grism.
For the  $g'r'i'$ photometry, we took one 2 s exposure and three 10 s
exposures dithered by $\sim 20\arcsec$ for each band.
For the photometric calibration 
the SDSS standard star BD +25$\arcdeg$4655\citep{smith02} was observed
nearly at the same airmass as target (difference $\lesssim$ 0.16).
Twilight images were also taken for the flat-fielding.
The limiting magnitudes were $g'$ $\sim$ 20.9, $r'$ $\sim$ 20.6, and $i'$
$\sim$ 20.0 for 10 $\sigma$ detections.

\begin{figure}
  \begin{center}
    \FigureFile(80mm,80mm){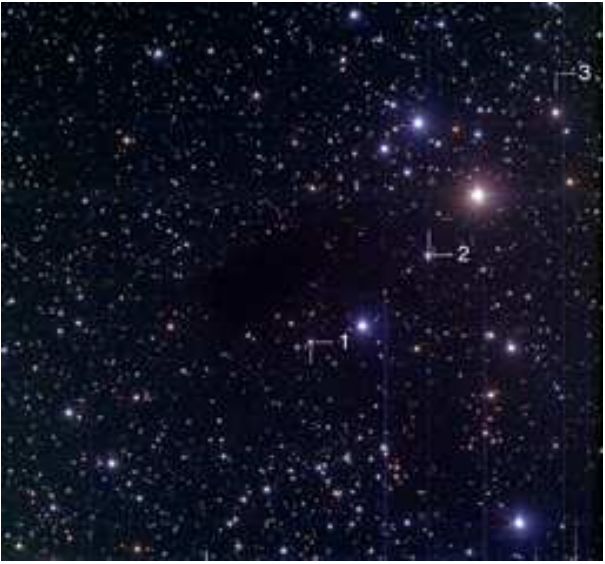}
  \end{center}
  \caption{Three-color composite image of L1014. The $g'r'i'$ data are represented here as blue, 
  green, and red, respectively. The area of the image is 11$\farcm$5 $\times$ 11$\farcm$5. 
  The three H$\alpha$ emission line stars are indicated with numbers 1 to 3. North is at the top, 
  east to the left.}\label{fig1}
\end{figure}

We extracted about 2000 spectra from the slit-less spectroscopy image and 
examined them for the presence of H$\alpha$ emission.
As a result, we found three H$\alpha$ emission line stars (figure~\ref{fig1}).
From the measurements of equivalent widths (EWs), we identified one candidate (star-1) 
as weak-line T Tauri star (WTTS) and two candidates (star-2 and 3) as classical 
T Tauri stars (CTTS).  Star-1 and 2 were previously reported as H$\alpha$ emission line 
stars toward B362, but their EWs were not measured  \citep{OH1983}.
The coordinates, EWs, magnitudes and colors of these stars are shown in
table \ref{prop-of-str}.

\begin{table*}
  \caption{Properties of T Tauri Star Candidates}\label{prop-of-str}
  \begin{center}
      \begin{tabular}{ccccccccc}
     \hline\hline
     Star & RA (J2000) & DEC (J2000) & EW & $g'$ & $r'$ & $i'$ & $g'-i'$ & Candidate\\
       &  & &  [\AA]  &  [mag]  &  [mag]  & [mag] & [mag] &Type\\
\hline
1 & 21$^h$23$^m$58.51$^s$ & +49$\arcdeg$58$\arcmin$09$\arcsec$ & 5.4 & 15.02$\pm$0.03 & 14.20$\pm$0.02 & 13.46$\pm$0.01 & 1.56 &  WTTS \\
2 & 21$^h$23$^m$44.72$^s$ & +49$\arcdeg$59$\arcmin$48$\arcsec$ & 21.2 & 13.74$\pm$0.03 & 13.11$\pm$0.02 & 12.74$\pm$0.01 & 1.00 & CTTS \\
3 & 21$^h$23$^m$29.69$^s$ & +50$\arcdeg$03$\arcmin$13$\arcsec$ & 14.9 & 18.54$\pm$0.04 & 16.94$\pm$0.02 & 15.99$\pm$0.01 & 2.55 & CTTS\\
 \hline
\multicolumn{4}{@{}l@{}}{\hbox to 0pt{\parbox{85mm}{\footnotesize Note. The magnitudes and colors are derived without a color correction on the assumption that the photometry system used here is 
well matched to the SDSS standard system \citep{smith02}. }\hss}}    
    \end{tabular}
  \end{center}
\end{table*}

The observed colors of star-1, 2, and 3 correspond to 
types K6--K7, K2--K3, and M3--M4, respectively.
For this comparison, we transformed the colors of main-sequence stars
bluer than M0 from the $UBVR_{\rm c}I_{\rm c}$ system \citep{KH95} 
to the $u'g'r'i'z'$ system with the equations in \citet{smith02}.
For stars whose colors are between M0 and M5, we adopted a color
transformation estimated from a comparison between our $g'r'i'$ photometry
of NGC 2264 and the $VR_{\rm c}I_{\rm c}$ photometry of \citet{DS05}.

To classify the spectral types of the T Tauri star candidates, 
we performed slit spectroscopy using WFGS2 during 2005 November 25--28 UT.
We obtained low-dispersion spectra of star-1 and 2 with a 300 line mm$^{-1}$
grism and a 0$\farcs$7 slit.
The spectra covered 426--810 nm with a resolving power of $R \sim$ 820 at 650 nm.
The total integration times were 2350 s for star-1 and 1200 s for star-2.
We also obtained a higher-dispersion spectrum of star-2 using a Volume-Phase
Holographic (VPH) grism \citep{ebizuka03} and  a 0$\farcs$7 slit.
The VPH grism spectrum covered 586--742 nm with a resolving power of 
$R \sim$ 4200 at 664 nm.  The total integration time was 2940 s.

\begin{figure}
  \begin{center}
    \FigureFile(80mm,80mm){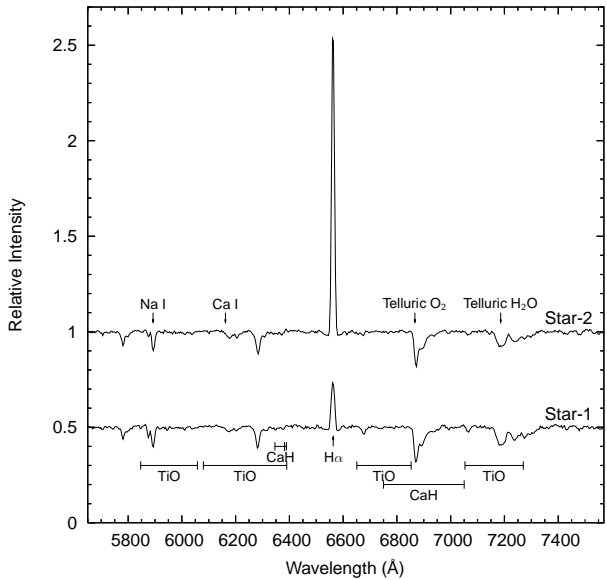}
  \end{center}
  \caption{Low-dispersion spectra of star-1 and 2. The normalized spectra are shown. 
  The positions of spectral features of the typical K-type stars and telluric absorption 
  bands listed in \citet{AS95} are also shown for comparison.}\label{fig2}
\end{figure}

\begin{figure}
  \begin{center}
    \FigureFile(80mm,80mm){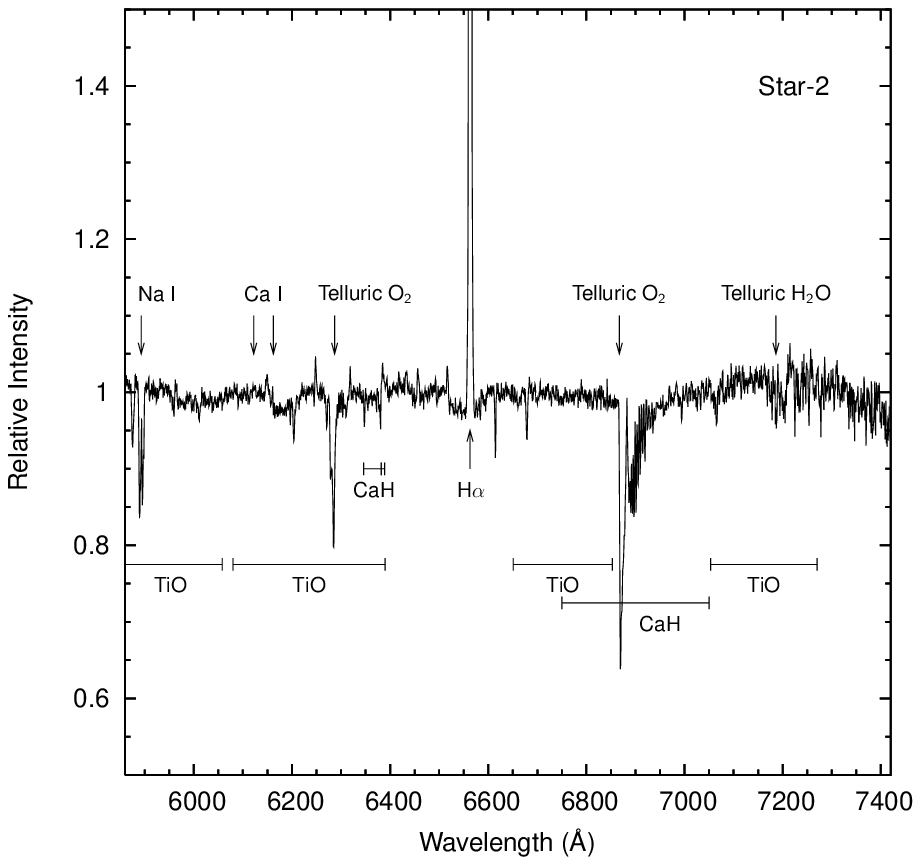}
  \end{center}
  \caption{Higher-dispersion spectrum of star-2. The normalized spectrum is shown. 
  The positions of spectral features of the typical K-type stars and telluric absorption 
  bands listed in \citet{AS95} are also showed for comparison.}\label{fig3}
\end{figure}

We classified the stars by comparing the spectra with the spectral atlas of
\citet{AS95} after normalizing each observed spectrum by fitting the
continuum with a spline function.
The features of the absorption lines in the low-dispersion spectra
of star-1 and 2 are almost identical, closely resembling a K3V spectrum
(figure~\ref{fig2}). However the H$\alpha$ emission line intensities differ.
In addition, three absorption lines of CaH (6346, 6382, and 6389 \AA), which are
features of mid K through M type stars, are seen in the higher-dispersion 
spectrum of star-2 (figure~\ref{fig3}).  Furthermore, 
a low-dispersion spectrum over 420--700 nm of star-1  was obtained
with the 65-cm telescope of the Gunma Astronomical Observatory 
on 2005 December 21 UT,  and it is consistent with the spectrum of 
reddened K2V to K4V star with a visual extinction of $\sim$ 0.8--1.5 mag.
Thus, we classified both star-1 and 2 as K3 type.
The error of classification is one or two in the subclass.

\section{Discussion}

\subsection{Are the H$\alpha$ emission line stars T Tauri stars?}

We detected three H$\alpha$ emission line stars near the small dense core.
These emission line stars are likely to be T Tauri stars because
of their proximity to the molecular cloud.  
However, some other possibilities, including those noted by 
\citet{DS05},  should also be considered: 

\begin{itemize}
\item[1)] classical Be stars,
\item[2)] Herbig Ae/Be stars,
\item[3)] dMe stars (EW $<$ 10 \AA, late K through M types),
\item[4)] chromospherically active giants, and 
\item[5)] RS CVn binaries, cataclysmic variables and symbiotic variables (EW $>$ 10 \AA).
\end{itemize} 

For star-1, possibilities  1)--3) are eliminated because the classified spectral type is 
not B or A and is earlier than late K types. 
Possibility 5) is unlikely because such stars usually have an H$\alpha$ emission line 
EW larger than 10 \AA.
Given star-1's observed  $g'-i'$ color of 1.56 mag, and assuming an intrinsic color 
$\sim$ 1.05 mag from the color of a K3V star in the $u'g'r'i'z'$ system described in  \S 2, 
we obtain a color excess of $\sim$ 0.51 mag.
We estimate the visual extinction of star-1 to be  $\sim$ 0.88 from this color excess 
and $A_V = E(g'-i') / 0.58$ \citep{FM03}.
On the other hand, the total extinction of the cloud toward star-1 is
about 2--3 mag from the C$^{18}$O and visual extinction maps of \citet{crapsi05}.
Therefore, star-1 is probably embedded in the L1014 core envelope, 
eliminating possibility 4).
We examined the Spitzer archival data of L1014, obtained by the c2d legacy program 
\citep{evans03}.  We identified star-1 with object SSTc2d J212358.5+495809. 
Its fluxes at 3.6, 4.5, 5.8, 8.0 and 24.0 $\micron$   are 11.2 $\pm$ 0.473, 9.97 $\pm$ 0.121, 
6.89 $\pm$ 0.258, 4.23 $\pm$ 0.0379, and 1.42 $\pm$ 0.0759 mJy, respectively.  
These fluxes give IRAC colors of [3.6]$-$[4.5] $\sim$ 0.36 and [5.8]$-$[8.0] $\sim$ 0.11, 
placing it outside the Class II region on the IRAC color-color diagram 
(see figure 4 of \cite{allen04}).  
These colors and the detection at  24.0 $\micron$ support that star-1 is  
WTTS with a thin disk or a disk having a large inner hole.
Many stars are detected in optical wavelengths as can be seen in figure~\ref{fig1}, but only 
a small number, including L1014-IRS, are detected on the 24.0 $\micron$  image, 
suggesting a probable association of circumstellar material with them.

For star-2, possibilities 1)--3) are eliminated for the same reasons as star-1.
Star-2's observed color is almost the same as that expected from its spectral type, 
whereas the C$^{18}$O map \citep{crapsi05} gives an extinction of $\sim$ 0.5 mag 
at its location.  We conclude that star-2 is located on the outskirts of the core, 
eliminating possibilities 4) and 5). 
Star-2 is outside the Spitzer IRAC images and is not cataloged, but it is detected 
at the edge of the 24.0 $\micron$ image and its 24.0 $\micron$ flux is larger than 
that of star-1.  This also suggests that star-2 is CTTS.

Since star-3's EW is larger than 10 \AA,  it is unlikely to be 3).
Although the extinction of star-3 is unclear because its spectral type is
unknown, the upper limit of extinction can be estimated to be $\sim$ 1.3 mag
from the $A_V$ map of \citet{dobashi05}\footnote{http://darkclouds.term.jp/}.
Even if the upper limit is adopted for the extinction, the color of star-3
is still red (late K type) relative to A or B type stars. 
Therefore, star-3 can not be 1) or 2).
We note that the lower limit of cloud extinction toward star-3 can be estimated 
to be $\sim$ 0.1 mag from the $^{13}$CO map of \citet{crapsi05}. 
Possibilities 4) and 5) remain, however these types of stars are extremely rare.
For example, \citet{Neuhauser97} classified only a few candidates for RS CVn binaries 
out of a $\sim$ 300 square degree area in the south of the Taurus molecular clouds
based on their X-ray and optical studies.

The CO and $A_V$ maps have relatively low spatial resolution, and provide average extinctions 
toward these stars.  It would be possible that the cloud is clumpy, and 
that these stars are located toward a part of the cloud which is slightly thinner than the average.  
However, the reverse would be also possible.  Moreover, no major molecular clouds are 
identified between L1014 and the Perseus arm \citep{dame01}.   
Thus their association with L1014 is very likely since T Tauri stars are 
usually found near molecular clouds, and most of the CO emission probably comes 
from the L1014 region, although it is difficult to definitely eliminate 
the possibility that they are background sources.

\subsection{Color-Magnitude Diagram}

We constructed the color-magnitude diagram ($i'$ vs. $g'-i'$) for L1014 to
estimate ages and masses of the T Tauri star candidates.
Although it is not yet clear whether star-3 is associated with the L1014 core or not,
here we assume that all the three stars are associated with the cloud.
We adopted the isochrones and the evolutionary tracks of \citet{PS99}.

\begin{figure}
  \begin{center}
    \FigureFile(80mm,80mm){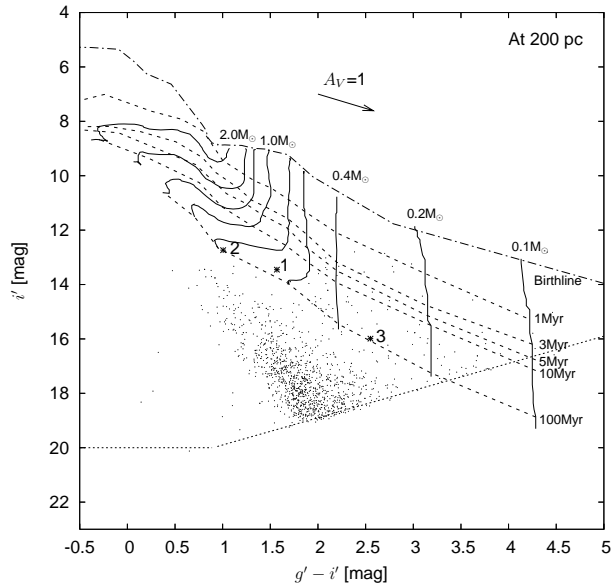}
  \end{center}
  \caption{Color-magnitude diagram of L1014 for distance of 200 pc. Star-1, star-2, and star-3 are 
  marked by numbered asterisk ($\ast$)  without extinction corrections.
  The isochrones (1, 3, 5, 10, and 100 Myr)  and evolutionary tracks (0.1, 0.2, 0.4, 0.6, 0.8, 1.0, 1.2, 1.5, 
  and 2.0 $M_\odot$) of \citet{PS99} are superposed. 
Since the dashed lines of isochrones are nearly parallel to the reddening vector, 
an extinction correction does not change an age estimate of each star significantly 
if the extinction is not so large.
  }\label{fig4}
\end{figure}

First, we assumed a distance of 200 pc for L1014 \citep{LM99, yng04} to
construct the color-magnitude diagram (figure~\ref{fig4}).
The diagram indicates ages of about 100 Myr.
This is not consistent with the typical age, $\lesssim$ 10 Myr, 
of T Tauri stars (e.g., \cite{PS04}).
Thus, our observations are incompatible with a distance of 200 pc.

\begin{figure}
  \begin{center}
    \FigureFile(80mm,80mm){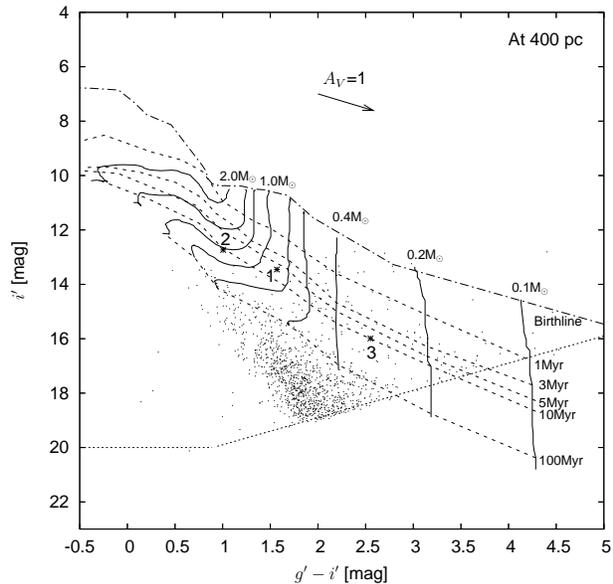}
  \end{center}
  \caption{Color-magnitude diagram of L1014 for 400 pc. Star-1, star-2 and star-3 are 
  represented by  numbered asterisk ($\ast$) without extinction corrections.
  The isochrones and evolutionary tracks of \citet{PS99} are also plotted in the diagram.
  For mass estimations of these stars, the assumed extinctions were 
  applied  (see Section 3.2).
  }\label{fig5}
\end{figure}

Second, we assumed a distance of 400 pc for L1014, as suspected by \citet{yng04}.
Figure~\ref{fig5} shows the color-magnitude diagram for this distance and
indicates that the stars' ages lie between $\sim$ 5 Myr and 10 Myr.  
This is consistent with the typical age of $\lesssim$ 10 Myr.
Since these stars are not embedded in the dense core
but located toward the envelope or more outer region, 
they could not be so young and probably have ages of $\gtsim$ 1 Myr.
Thus $\sim$ 1--10 Myr is plausible as their age range.  Adopting this range, 
$\sim$ 400--900 pc seems to be more reasonable than 200 pc.

Using the color-magnitude diagram at a distance of $\sim$400--900 pc
and applying an extinction correction for each star, 
we estimate masses of $\sim$ 1.2--1.8 $M_\odot$, $\sim$ 1.2--1.8 $M_\odot$, 
and 0.3--0.5 $M_\odot$ for star-1, 2, and 3, respectively.
Extinctions of 0.88 and 0.1--1.3 were adopted for star-1 and 3, and no
extinction was assumed for star-2.

The locations of the three stars in the color-magnitude diagram suggest that 
these stars have almost the same age. This suggests 
that star-3 is at the same distance as the other two
associated with the L1014 cloud and that it might also be associated with
the cloud.

\section{Conclusions}

We identified three T Tauri stars toward the L1014 core by slit-less spectroscopy,
$g'r'i'$ photometry and slit spectroscopy.
One of them is a WTTS and the others are CTTS.
The estimation of the extinction and the ages of T Tauri stars suggests that they are
associated with L1014. This indicates that star formation took place previously 
in L1014's  progenitor cloud.   

An examination of the color-magnitude diagram suggests 
that the distance to L1014 is $\sim$ 400--900 pc. 
Since the previously assumed distance of 200 pc was based on the kinematic distance, 
our estimated distance is plausible
if these stars are associated with L1014.

The luminosity of the very low mass protostar formed in the L1014 core is reported
to be $\sim$ 0.09 $L\odot$, for a distance of 200 pc \citep{yng04}.
Assuming a distance of 400--900 pc, the mass estimate of the protostellar object 
(L1014-IRS) significantly increases.

\bigskip

We are grateful to the staff of the UH 2.2-m telescope for supporting our
WFGS2 observations.
We also appreciate the great support of the Subaru Telescope.
Use of the UH 2.2-m telescope for the observations is supported by the National
Astronomical Observatory of Japan. We thank  Joel F. Koerwer for proof reading.
A. M. thanks the Daiko Foundation for their financial support.
WFGS2 was supported by the Daiko Foundation and by the Sumitomo Foundation.
This instrument was also supported by Grants-in-Aid from the Ministry of
Education, Culture, Sports, Science and Technology (14540228 and 15540238).
This work was financially supported in part by Grants-in-Aid from the Ministry
of Education, Culture, Sports, Science and Technology (16740121 and 17039011).


\end{document}